\documentclass[12pt]{article}

\newcommand{\dZ}{Z \! \! \! Z}
\begin{document}

\title{Invariant Submanifolds of Darboux-KP Chain and Extension of the
Discrete KP Hierarchy}

\author{A.K. Svinin}

\maketitle

\begin{abstract}
Invariant submanifolds of the so-called Darboux-KP chain \cite{ma1} are
investigated. It is shown that restriction of dynamics on some class of
invariant submanifolds yields the extension of the discrete KP hierarchy
while the intersections of the latter lead to Lax pairs for a broad class
of differential-difference systems with finite number of fields. Some attention is given to investigation
of self-similar reductions. It is shown that self-similar ansatzes lead to
purely discrete equations with dependence on some number of parameters together
with equations governing deformations with respect to these parameters. Some
examples are provided. In particular it is shown that well known discrete
first Painlev\'e equation (dPI) corresponds to Volterra lattice hierarchy.
It is written down equations which naturally generalize dPI in the sense
that they have first Painlev\'e transcedent in continuous limit.
\end{abstract}
\section{Introduction}

In the work \cite{ma1}, was introduced the notion of Darboux-KP (DKP) hierarchy
realizing the fundamental concept of Darboux covering adapted for the flows of
KP hierarchy. This notion proved to be instrumental for investigation of
invariant submanifolds of the KP hierarchy and also for other aims.
In fact the DKP hierarchy represents two copies (solutions) of KP hierarchy
glued together by Darboux map. Iteration of Darboux map, in both directions
yields DKP chain

In the article \cite{sv1} we constructed two-parameter class of invariant
submanifolds of the DKP chain phase space ${\cal S}_l^n$.
In particular case $n=1$, these submanifolds were found in \cite{ma1}.
It was shown there that on ${\cal S}_0^1$ the DKP chain is reduced to
discrete KP hierarchy \cite{ue1}, \cite{ue2}.

This work is concerned with investigation of DKP chain phase space invariant
submanifolds. In Section 3 we show that restriction of the DKP chain on
${\cal S}_0^n$ leads to well-defined Lax equations which can be written down
in explicit form. The collection of all the flows of the DKP chain
restricted on the class of submanifolds  $\{{\cal S}_0^n : n\geq 1\}$
naturally form the extension of the discrete KP hierarchy
Restriction of the DKP chain on intersections like ${\cal S}_{n,r,l}=
{\cal S}_0^n\cap{\cal S}_{l-1}^{ln-r}$, in turn, gives Lax pairs for
differential-difference systems with finite number of fields.
These lattice govern Darbox transformations for (restricted) KP hierarchy
Lax operators.

To  show efficiency of this approach for constructing of integrable lattices,
in Section 4, we provide the reader by some examples of integrable
lattices, which can be found in the literature and also we construct
the class of one-component lattices which naturally includes Bogoyavlenskii
ones. We show in this paper that investigation of invariant submanifolds
of DKP chain allows to construct Miura transformations between lattices
under consideration.

In Section 5 we investigate the solutions of lattice hierarchy invariant
with respect to dilatations. It is shown that suitable ansatzes
lead to purely discrete equations depending on some collections of parameters
together with equations describing deformations of these parameters.
Typical example is dPI corresponding to Volterra lattice hierarchy.
In particular, we recover the fact that dPI describes Schlezinger transformation
of the PIV equation. We investigate class of the discrete equations
corresponding to Bogoyavlenskii lattices. All these systems pass singularity
confinement test provided some condition on the constants entering these
systems. It is shown that this condition is more general than in integrable
case. It is shown that all these systems turn into PI in continuous limit.

\section{Darboux-KP hierarchy}

{\bf 2.1. KP hierarchy.}

First of all let us recall the formalism of the KP hierarchy along the lines
proposed in \cite{ca2},
\cite{ca1}, \cite{fa1}. One considers the space of Laurent series (currents)
of the form
\[
H^{(0)} = 1,\;\;
H^{(p)} = z^p + \sum_{l\geq 1}H_l^pz^{-l}.
\]
The point of the phase space is defined by semi-infinite matrix
$(H_l^p)_{l\geq 1, p\geq 1}$. With each point one associates linear span
${\cal H}_{+} = <1, H^{(1)}, H^{(2)},...>$ in the space of Laurent series
\[
H= \left\{\sum_{-\infty\ll k<\infty}l_kz^{-k}\right\} =
{\cal H}_{+}\oplus H_{-}.
\]
It is evident that $H_{-} = <z^{-1}, z^{-2},...>$. One considers invariance
relation
\begin{equation}
(\partial _p + H^{(p)}){\cal H}_{+}\subset {\cal H}_{+},\;
p\geq 1
\label{inv}
\end{equation}
which one writes in explicit form
\begin{equation}
\partial_pH^{(k)} = H^{(k+p)} - H^{(k)}H^{(p)} + \sum_{s=1}^pH_s^kH^{(p-s)}
+ \sum_{s=1}^kH_s^pH^{(k-s)}.
\label{CS}
\end{equation}
These equations are called Central System (CS). It is obvious exactness property
$\partial_pH^{(k)} = \partial_kH^{(p)}$. Using the latter it is easy to prove
the commutativity of SC flows.

The passage to KP hierarchy needs spatialization of some evolution parameter,
namely $t_1=x$.  Putting in (\ref{CS}) $p=1$ one obtains
\begin{equation}
(\partial + h)H^{(k)} = H^{(k+1)} + H_1^k + h_2H^{(k-1)} + ... + h_{k+1}
\label{CS1}
\end{equation}
where $h = z+ h_2z^{-1} + h_3z^{-2} + ... \equiv H^{(1)}$.
Using the relation (\ref{CS1}) one can easily show that $H^{(k)}$'s are
expressed as linear combinations on Fa\`a di Bruno
differential polynomials (iterations) which are defined by the following
recurrence relations:
\[
h^{(k+1)} = (\partial + h)h^{(k)},\;\;
h^{(1)} = h,\;\;
k\geq 1,
\]
and under these circumstances one has
${\cal H}_{+} = <1, h^{(1)}, h^{(2)}, ...>$. One writes the formula of passage
from the basis $\{h^{(k)}\}$ to basis $\{H^{(p)}\}$ as
\[
H^{(p)} = h^{(p)} + \sum_{k=0}^{p-2}r_k^p[h]h^{(k)}.
\]

On this stage one has passage from the phase space of semi-infinite matrices
to KP hierarchy one whose points are defined by infinite collections
of (smooth) functions$\{h_k(x),\; k\geq 2\}$. Using the exactness property
one can write
\begin{equation}
\partial_ph = \partial H^{(p)}.
\label{KP}
\end{equation}
Since $H^{(p)} = z^p + {\cal O}(z^{-1})$ then $H^{(p)} = \pi_{+}(z^p)$ where
$\pi_{+}$ denotes projection of arbitrary Laurent series into ${\cal H}_{+}$.

The relation (\ref{KP}) defines infinite number of evolution equations in the
form of local conservation laws. As is known, these equations are entirely
equivalent to KP hierarchy while $h(z)$ is interpreted as generating function
of Hamiltonian densities. The passage from Laurent series to the algebra
of pseudodifferential operators governed by the rule $\phi(h^{(k)}) =
\partial^k$ which is extended by linearity on the whole space $H$.
Since negative powers of $\partial$ enter Lax KP operator, one needs
to define negative Faa di Bruno iterations with the help of the relations
\[
(\partial + h)h^{(-1)} = 1,\;\;
(\partial + h)h^{(-2)} = h^{(-1)}\;\;
\]
and so on.
Expressing $z$ as linear combination on Faa di Bruno iterations one defines
Lax operator ${\cal Q} = \partial + u_2\partial^{-1} +
u_3\partial^{-2} + ... $ as ${\cal Q} = \phi(z)$. It is easy from this,
to win
\[
u_2 = -h_2,\;\;
u_3 = -h_3,\;\;
u_4 = -h_4 - h_2^2,\;\;
u_5 = -h_5 - 3h_2h_3 + h_2h_{2x}
\]
and so on. It can be seen, that the relation (\ref{KP}) is equivalent to
Lax equation
$\partial_p{\cal Q} = [{\cal Q}^p_{+}, {\cal Q}]$.

The relations connecting formal Baker-Akhiezer function
$\psi = (1 + \sum_{k\geq 1}w_kz^{-k})
\exp\sum_{p\geq 1}t_pz^p$ with Fa\`a di Bruno iterations and currents are
as follows:
\begin{equation}
h^{(k)} = \frac{\partial^k\psi}{\psi},\;\;
H^{(p)} = \frac{\partial_p\psi}{\psi}.
\label{flow}
\end{equation}

{\sc Remark 1.}
The representation of the KP hierarchy in the form of local conservation
laws (\ref{KP}) is equivalent to Wilson's theorem \cite{wi1}.
Noncommutative variant of this theorem can be found in \cite{ku2}.
Central System in the form (\ref{CS}) comes back to Cherednik's results  
(cf. \cite{ku2}).

\noindent
{\bf 2.2. DKP hierarchy.}

In the article \cite{ma1} was defined DKP hierarchy
\begin{equation}
\begin{array}{l}
\partial_p h = \partial H^{(p)},  \\
\partial_p a = a(\tilde{H}^{(p)} - H^{(p)}).
\end{array}
\label{DKP}
\end{equation}
Here laurent series $a = z + \sum_{k\geq 0}a_{k+1}z^{-k}$ differs from $h$
by the presence of zero power of $z$. ``New'' currents $\tilde{H}^{(p)}$
are calculated at the point $\tilde{h} = h + a_x/a$.

The equations (\ref{DKP}) realize the concept of Darboux covering
adapted to dynamical systems of KP hierarchy. Indeed, it is easy to show that
having some solution $(h, a)$ of the system (\ref{DKP}) one can construct new
one with the help of Darboux map $\sigma(h, a) = h + a_x/a$. It was shown
in the work \cite{ma1} that this construction allows enough simply describe
many notions using in KP hierarchy theory like Krichever's rational reductions,
\cite{kr1}, Darboux and Miura transformations \cite{ko1}, \cite{ji1},
discrete analog of KP hierarchy \cite{ue1}, \cite{ue2}.

One can look at (\ref{DKP}) from another viewpoint. Given any pair of
KP hierarchy solutions $(h, \tilde{h})$ or equivalently
$(\psi, \tilde{\psi})$ one uniquely defines $a$. It is easy observe
that substitution of $a = z\tilde{\psi}/\psi$ in (\ref{DKP}), taking into
account (\ref{flow}), turn these equations into identities.
In particular the pair $(h, h)$ is suitable for trivial solution $a=z$.
From this standpoint the system (\ref{DKP}) do not seem to be useful.
It becomes really  informative after imposing some restrictions compatible
with this system. As is known \cite{ma1}, the condition
$z^la\in {\cal H}_{+}$ at $l\geq -1$ defines invariant submanifold
${\cal S}_l$ for (\ref{DKP}). It is important to observe that in these
circumstances KP hierarchy rests to be nonrestricted, but
the mapping $h\rightarrow\tilde{h}$ specifies.
For instance, on ${\cal S}_{-1}$ one has $a=z$ and correspondingly $\tilde{h}=h$.
On ${\cal S}_0$, in turn we have $a=h+a_1$ and
\[
\tilde{h} = \frac{a_x + a^2 -a_1a}{a} = \frac{a^{(2)} - a_1a^{(1)}}{a}.
\]
In terms of wave functions, on ${\cal S}_0$ we have $z\tilde{\psi} = (\partial +
a_1)\psi\equiv H\psi$ (elementary Darboux transformation).
This transforms is controlled by the function $a_1$. It stands to reason that
it can not be arbitrary but must satisfy some equations, namely
\cite{ma1}:
\begin{equation}
\partial_pa_1 + \partial\left((-a_1)^{(p)} +
\sum_{k=0}^{p-2}r_k^p[h](-a_1)^{(k)}\right) = 0.
\label{a1}
\end{equation}
Here $(-a_1)^{(k)}$ are corresponding Fa\`a di Bruno iterates
\[
(-a_1)^{(0)} = 1,\;\;
(-a_1)^{(1)} = -a_1,\;\;
(-a_1)^{(2)} = -a_{1x} + a_1^2
\]
and so on. Equation (\ref{a1}) can be exactly linearizable by ansatze
$a_1 = - \Phi_x/\Phi$. In a result, one obtains $\partial_p\Phi =
{\cal Q}_{+}^p(\Phi)$.  $\tau$-function transforms especially simply:
$\tilde{\tau} = \Phi\tau$ \cite{ch1}.

Constrained KP hierarchies  (Krichever's rational reductions) arises
as intersections ${\cal S}_l\cap {\cal S}_{l+r}$. In this case $r$-th power
of Lax operator turns out to be expressible as a ratio of two purely differential
operators 
\begin{equation}
{\cal Q}^r= P_{l+1}^{-1}Q_{l+r+1}
\label{rat}
\end{equation}
(cf. \cite{di2}).
In particular, the case $l=-1$ is suitable for reductions to Gelfand-Dickey
hierarchies \cite{ge1}.

%----------------------------------------------------------------------------

\section{Extension of the discrete KP hierarchy}

\noindent
{\bf 3.1. DKP chain and its invariant submanifolds.}

Applying Darboux iteration to some fixed solution of DKP hierarchy
infinitely many times we obtain infinite collection of Laurent series
$\{h(i), a(i) : i\in\dZ\}$ satisfying the system
\begin{equation}
\begin{array}{l}
\partial_p h(i) = \partial H^{(p)}(i),  \\
\partial_p a(i) = a(i)(H^{(p)}(i+1) - H^{(p)}(i)).
\end{array}
\label{DKPi}
\end{equation}

{\sc Remark 2.}
To each copy of KP hierarchy corresponds copy of CS parametrized by some
value of $i$, so ${\cal H}_{+}(i) = <1, H^{(1)}(i), H^{(2)}(i),...>$.

As in \cite{sv1} the system (\ref{DKPi}) is reffered to as DKP chain .
In the work \cite{sv1} was found denumerable class of submanyfolds
invariant with respect to the flows (\ref{DKPi}).

{\sc Theorem.}
{\it Submanifold ${\cal S}_l^n$ defined by the condition
\begin{equation}
z^{l-n+1}a^{[n]}(i)\in{\cal H}_{+}(i),\;\;
\forall i\in\dZ
\label{th}
\end{equation}
with $n\in\dZ^{*}\equiv\dZ / \{0\}$ is tangent with respect to DKP chain
flows.
}

Here $a^{[k]}(i)$ are ``discrete'' Fa\`a di Bruno iterates defined
by recurrence relations $a^{[k+1]}(i) = a(i)a^{[k]}(i+1)$ with
$a^{[0]}(i)\equiv 1$. For $k>0$,
\[
a^{[k]}(i) = a(i)a(i+1)...a(i+k-1)
\]
and for $k<0$
\[
a^{[k]}(i) = a^{-1}(i-1)a^{-1}(i-2)...a^{-1}(i-|k|).
\]

{\sc Proposition 1.}
{\it By virtue of the second equation in (\ref{DKPi})
\begin{equation}
\partial_p a^{[k]}(i) = a^{[k]}(i)(H^{(p)}(i+k) - H^{(p)}(i)).
\label{DKPii}
\end{equation}
}

{\sc Proof.} For $k>0$ we have
\[
\partial_pa^{[k]}(i) =
\partial_p[a(i)a(i+1)...a(i+k-1)] =
\]
\[
= \sum_{s=1}^k
a(i)...a(i+s-1)\{H^{(p)}(i+s) - H^{(p)}(i+s-1)\}a(i+s)...a(i+k-1) =
\]
\[
= a^{[k]}(i)(H^{(p)}(i+k) - H^{(p)}(i)).
\]
Analogous calculations are performed for negative $k$.

{\sc Corollary.}
{\it Define, for any integer $k\neq 0$, $h = h(i),\; \tilde{h} = h(i+k),\;
a = z^{1-k}a^{[k]}(i)$ then, by virtue of (\ref{DKPi}), the triple
$(h, \tilde{h}, a)$ is a solution of DKP hierarchy.
}

{\sc Proposition 2.}
{\it The following chain of inclusions is valid:
\begin{equation}
{\cal S}_l^n\subset {\cal S}_{2l+1}^{2n}\subset...\subset {\cal S}_{kl+k-1}^{kn}
\subset...
\label{inclusion}
\end{equation}
}

{\sc Proof.} Let us show that on ${\cal S}_l^n$
\begin{equation}
z^{l-n+1}a^{[n]}(i){\cal H}_{+}(i+n)\subset{\cal H}_{+}(i).
\label{inclusion1}
\end{equation}
As can be checked the formula (\ref{inclusion1}) follows from the relation
\begin{equation}
z^{l-n+1}a^{[n]}(i)h^{(k)}(i+n) =
(\partial + h(i))^k z^{l-n+1}a^{[n]}(i),\;\;
\forall k\geq 0
\label{inclusion2}
\end{equation}
together with invariance relation (\ref{inv}). In turn,
(\ref{inclusion2}) can be proved by induction. For $k=0$,
(\ref{inclusion2}) is obvious. Let us suppose that this relation is
valid for some value of $k$, then by virtue of (\ref{DKPii}) we have
\[
(\partial + h(i))^{k+1}z^{l-n+1}a^{[n]}(i)
= z^{l-n+1}a^{[n]}(i)(h(i+n) -
h(i))h^{(k)}(i+n) +
\]
\[
+ z^{l-n+1}a^{[n]}(i)\partial h^{(k)}(i+n) +
z^{l-n+1}a^{[n]}(i)h(i)h^{(k)}(i+n) =
z^{l-n+1}a^{[n]}(i)h^{(k+1)}(i+n).
\]
Then to show inclusions (\ref{inclusion}) one needs to use
(\ref{inclusion1}) and easily checked formula
\begin{equation}
z^{l-n+1}a^{[n]}(i)\cdot z^{p(l-n+1)}a^{[pn]}(i+n) =
z^{(p+1)(l-n+1)}a^{[(p+1)n]}(i).
\label{aux}
\end{equation}
From the condition $z^{l-n+1}a^{[n]}(i)\in{\cal H}_{+}(i)$ and relations
(\ref{inclusion1}) and (\ref{aux}) we obtain
$z^{2(l-n+1)}a^{[2n]}(i)\in{\cal H}_{+}(i)$ and so on.

{\sc Definition.}
Suppose that the solution of DKP chain is in ${\cal S}_l^n$ and there is not
invariant submanifold defined by the condition (\ref{th}) which: 1) is in
${\cal S}_l^n$; 2) contains given solution then one says that
${\cal S}_l^n$ is origin of the chain of inclusions for this solution.

{\sc Remark 3.} The reasonings used in proof of Proposition 2
can be found in \cite{sv1} but this proposition was not exhibited there.

{\sc Remark 4.} In the article \cite{sv1} we suppose that $n\geq 1$ but
this is not necessary. From the viewpoint of constructing of integrable
lattices consideration of negative values of $n$ gives not something
new since with the help of invertible transformation  $g_{-1}$ (see below
(\ref{trans}) corresponding to inversion of the discrete parameter:
$i\rightarrow -i$, one always can pass to positive $n$. On the other hand
when considering intersections of invariant submanifolds there is a need
to use ${\cal S}_l^n$ with negative value of $n$.

The invariance property of ${\cal S}_l^1$ was exhibited in \cite{ma1}. In the
case ${\cal S}_0^1$ DKP chain is reduced to discrete KP hierarchy. In the next
section we investigate more general case when $n\geq 1$.

\noindent
{\bf 3.2. Extension of the discrete KP hierarchy.}

The relation defining ${\cal S}_0^n$ looks very simply
\[
h(i) = z^{1-n}a^{[n]}(i) - a_1^{[n]}(i) \equiv z^{1-n}a^{[n]}(i) -
\sum_{s=1}^na_1(i+s-1)
\]
that is here $h_k(i)$'s are uniquely expressed as polynomials of $a_k(i)$.
This means that motion equations on submanifolds ${\cal S}_0^n$ can be
rewritten only in terms of coordinates $a_k(i)$. More precisely, the equations
obtained below (\ref{dKP}) define projection of the DKP chain flows
from invariant submanifolds on affine hyperplane whose points are parametrized
by coordinates $a_k(i)$ ($a$-surface).

In what follows it will be convenient to define the set of variables
$\{q_k^{(n,r)}(i) \}$ as functions of coordinates $a$-surface with the help of the relation
\begin{equation}
z^r = a^{[r]}(i) + \sum_{k\geq 1}q_k^{(n,r)}(i)z^{k(n-1)}a^{[r-kn]}(i).
\label{REL}
\end{equation}
Since $z^{k(n-1)}a^{[r-kn]}(i)$ is a monic Laurent series of power $r-k$
then the formula (\ref{REL}) uniquely defines $q_k^{(n,r)}(i)$
as polynomials in $a_k(i)$. Remark that this expression has not
relation to any invariant submanifold, but it serves for determining
the map $\{a_k(i)\}\rightarrow
\{q_k^{(n,r)}(i)\}$ for some fixed $n$ and $r$.

Let us define, for each $n$, the set of ``discrete'' currents
$\{K_{(n)}^{[pn]}(i) : p\geq 1\}$ with the help of the relation
\[
K_{(n)}^{[pn]}(i) =
a^{[pn]}(i) + \sum_{k=1}^pq_k^{(n,pn)}(i)z^{k(n-1)}a^{[(p-k)n]}(i).
\]

{\sc Proposition 3.} \cite{sv1}
{\it On ${\cal S}_0^n$
\begin{equation}
z^{(1-n)p}K_{(n)}^{[pn]}(i) =
H^{(p)}(i).
\label{PR}
\end{equation}
}
The relation (\ref{PR}) allows to rewrite second equation in (\ref{DKPi})
in terms of coordinates $a_k(i)$
\begin{equation}
z^{p(n-1)}\partial_pa(i) = a(i)(K_{(n)}^{[pn]}(i+1)  -
K_{(n)}^{[pn]}(i)).
\label{dKP}
\end{equation}
In these circumstances, by virtue of invariance of ${\cal S}_0^n$
the first equation in (\ref{DKPi}) turn into identity and becomes
in a sense unnecessary. Let us observe that the equations
(\ref{dKP}) describing projections of the flows on different invariant
submanyfolds are also different what is quite natural. We attach to
evolution parameters corresponding to ${\cal S}_0^n$ the label:
$t_p = t^{(n)}_p$.

As was mentioned above, in the case $n=1$
the equations (\ref{dKP}) are equivalent to discrete KP hierarchy,
so it is quite natural to call the whole collection of the flows
describing by the system (\ref{dKP}) extended discrete KP (edKP) hierarchy.
But it is worth to remark that the flows corresponding to different labeles
do not be obliged to be commutative.

Our next tasks are to rewrite the system (\ref{dKP}) in the form of Lax equation;
to write down in its explicit form differential-difference equations
on variables $q_k^{(n,r)}(i)$ and to list possible reductions
of the system (\ref{dKP}) for different $n$.
As a result it gives the possibility to cover a broad class of the lattices
which admits Lax representation including such classical examples
as Volterra and Toda lattices.

\noindent
{\bf 3.3. Lax representation for edKP hierarchy.}

Let $\{\psi_i : i\in\dZ\}$ be the set of wave functions of KP hierarchy
corresponding to DKP chain (\ref{DKPi}). Let us define vector wave function
$\Psi$ with coordinates $\Psi_i = z^i\psi_i$. Then
\[
\Psi_i = z^i(1+\sum_{k\geq 1}w_k(i)z^{-k})\exp\sum_{p\geq 1}t_pz^p.
\]
It is obvious that the relationship between discrete Fa\'a di Bruno iterates
and $\Psi$ is defined by the following formulas:
\begin{equation}
a(i) = \frac{\Psi_{i+1}}{\Psi_i},\;\;
a^{[r]}(i) = \frac{\Psi_{i+r}}{\Psi_i}.
\label{discr}
\end{equation}
Using the relations (\ref{PR}) and (\ref{discr}), on ${\cal S}_0^{n}$,
one gets
\[
z^{p(n-1)}H^{(p)}(i) =
z^{p(n-1)}\frac{\partial_p^{(n)}\psi_i}{\psi_i} =
z^{p(n-1)}\frac{\partial_p^{(n)}\Psi_i}{\Psi_i} =
K_{(n)}^{[pn]}(i)  =
\]
\[
= \frac{\Psi_{i+pn}}{\Psi_i} +
\sum_{k=1}^pq_k^{(n,pn)}(i)z^{k(n-1)}\frac{\Psi_{i+(p-k)n}}{\Psi_i}
\]
or
\begin{equation}
z^{p(n-1)}\partial_p^{(n)}\Psi = (Q_{(n)}^{pn})_{+}\Psi,
\label{evolution}
\end{equation}
where
\[
(Q_{(n)}^{pn})_{+} \equiv
\Lambda^{pn} +
\sum_{k=1}^pq_k^{(n,pn)}(i)z^{k(n-1)}\Lambda^{(p-k)n};
\]
$\Lambda$ --- the shift operator acting by the rule $(\Lambda f)(i) = f(i+1)$.

The equation (\ref{evolution}) defines evolutions. To construct
Lax pair, one needs to find out eigenvalue problem. From (\ref{REL}) one gets
\[
z^r = \frac{\Psi_{i+r}}{\Psi_i} +
\sum_{k\geq 1}q_k^{(n,r)}(i)z^{k(n-1)}\frac{\Psi_{i+r-kn}}{\Psi_i}
\]
or
\begin{equation}
Q_{(n)}^r\Psi = z^r\Psi,
\label{eigenv.pr.}
\end{equation}
where
\[
Q_{(n)}^r \equiv
\Lambda^r +
\sum_{k\geq 1}q_k^{(n,r)}(i)z^{k(n-1)}\Lambda^{r-kn}.
\]
Remark once more that the relation (\ref{eigenv.pr.}) which is simply
(\ref{REL}) rewritting in terms of wave vector-function has not relation
to any invariant submanifold ${\cal S}_l^n$ (as opposed to (\ref{evolution})).

Consistency condition of linear auxiliary linear systems
(\ref{evolution}) and (\ref{eigenv.pr.}) is the equation
\begin{equation}
z^{p(n-1)}\partial_p^{(n)}Q_{(n)}^r =
[(Q_{(n)}^{pn})_{+}, Q_{(n)}^r].
\label{Lax}
\end{equation}
It can be written in its explicit form
\[
\partial_p^{(n)}q_k^{(n, r)}(i) = Q_{k,p}^{(n,r)}(i) = q_{k+p}^{(n, r)}(i+pn) -
q_{k+p}^{(n, r)}(i) +
\]
\begin{equation}
+ \sum_{s=1}^{p}q_s^{(n, pn)}(i)\cdot q_{k-s+p}^{(n,r)}(i+(p-s)n) -
\sum_{s=1}^{p}q_s^{(n, pn)}(i+r-(k-s+p)n)\cdot q_{k-s+p}^{(n,r)}(i).
\label{expl.form}
\end{equation}

{\sc Remark 5.}
The equation (\ref{Lax}) for $r=1$ is considered \cite{ku1}, \cite{ku2}
and is reffered to as gap KP hierarchy. More exactly,
for the operator $L = \Lambda +
\Lambda^{1-\Gamma}\circ q_0 + \Lambda^{1-2\Gamma}\circ q_1 + ...$ the equation
$\partial_pL = [L^{p\Gamma}_{+}, L]$ is considered.
For this equation, in \cite{ku2}, the problem of integrable discretization
of the flow is solved.

In what follows, we exhibit examples of differential-difference systems
resulting as different reductions from (\ref{expl.form}). When constructing
such systems it is important to take into account that the functions
$q_k^{(n,r)}(i)$ are not independent with respect to each other.
The relation (\ref{eigenv.pr.}) says that multiplication wave
vector-function by $z^r$ is equivalent to the action on it the operator
$Q_{(n)}^r$. Then
\[
z^{r_1+r_2}\Psi = Q_{(n)}^{r_1+r_2}\Psi = z^{r_1}Q_{(n)}^{r_2}\Psi
= Q_{(n)}^{r_2}Q_{(n)}^{r_1}\Psi
= z^{r_2}Q_{(n)}^{r_1}\Psi
= Q_{(n)}^{r_1}Q_{(n)}^{r_2}\Psi.
\]
From this it follows
\[
Q_{(n)}^{r_1+r_2} = Q_{(n)}^{r_1}Q_{(n)}^{r_2} =
Q_{(n)}^{r_2}Q_{(n)}^{r_1}
\]
or in more explicit form
\[
q_k^{(n,r_1+r_2)}(i) =
q_k^{(n,r_1)}(i) +
\sum_{s=1}^{k-1}q_s^{(n,r_1)}(i)q_{k-s}^{(n,r_2)}(i+r_1-sn)
+ q_k^{(n,r_2)}(i+r_1) =
\]
\begin{equation}
= q_k^{(n,r_2)}(i)
+ \sum_{s=1}^{k-1}q_s^{(n,r_2)}(i)q_{k-s}^{(n,r_1)}(i+r_2-sn)
+ q_k^{(n,r_1)}(i+r_2).
\label{r}
\end{equation}

Directly from (\ref{expl.form}) one sees that equations corresponding
to some fixed values of $n$ and $r$ admit reduction with the help
of the restriction
\begin{equation}
q_k^{(n,r)}(i)\equiv 0,\;\;
k>l,\;\; l\geq 1.
\label{reductions}
\end{equation}
From geometrical viewpoint these reductions, as was shown in \cite{sv1},
is relevant to intersections ${\cal S}_{n,r,l}\equiv
{\cal S}_0^n\cap {\cal S}_{l-1}^{ln-r}$.

%----------------------------------------------------------------------------

\section{Integrable lattices}

The aim of this section is to exhibit some examples of differential-difference
systems which can be derived from  (\ref{expl.form}) and (\ref{r}).
By its construction, they possess Lax pair representation (\ref{Lax})
with some reduced operator $Q_{(n)}^r$ and have direct relation to KP hierarchy
in the sense that they relate in some way the sequences of (restricted)
KP Lax operators.

\noindent
{\bf 4.1. Examples.}

First of all it should be notice recent work \cite{ci1} in which a broad class
of lattices having Lax pair was exhibited. But it can be shown that
the most part of these examples can be extracted from (\ref{Lax}).

On ${\cal S}_{n,r,l}$ we have
$
Q_{(n)}^r = \Lambda^r +
\sum_{k=1}^{l}z^{k(n-1)}q_k^{(n, r)}\Lambda^{r-kn}.
$
Denote $\lambda = z^{1-n}$ and $H(\lambda) = \lambda^lQ_{(n)}^r$. Then
\begin{equation}
H(\lambda) = \lambda^l\Lambda^r
+ \sum_{k=1}^{l}\lambda^{l-k}q_k^{(n, r)}\Lambda^{r-kn} =
\sum_{k=1}^{l}\lambda^kH_k,
\label{f}
\end{equation}
\[
\frac{\lambda^p}{\lambda^{Nl}}H^N(\lambda) = \lambda^p\Lambda^{pn} +
\sum_{k=1}^{\infty}\lambda^{p-k}q_k^{(n, r)}\Lambda^{(p-k)n}.
\]
Observe that Lax equation (\ref{Lax}), provided that $pn =Nr$ where $N$ is some
ineger, is rewritten in the following form:
\begin{equation}
\partial_p^{(n)}H(\lambda) = [\left(\lambda^{p-Nl}H^N(\lambda)\right)_{\infty},
H(\lambda)] =
- [\left(\lambda^{p-Nl}H^N(\lambda)\right)_{0},
H(\lambda)].
\label{ustinov}
\end{equation}
Here the subscripts $\infty$ and $0$  denote projections of Laurent series
(with matrix coefficients) on nonnegative and negative parts, respectively.
The equation (\ref{ustinov}) is basis for constructing integrable lattices,
Darboux transformations and soliton solutions in the work \cite{ci1}.
Unfortunately in this work, there is no indication how to construct
matrices $H_k$, while most part of examples presented in this work are suitable
for (\ref{f}).

The possibility to write down the (\ref{expl.form}) as the system
with finite number of fields in closed form, is that thanks to condition
$pn = Nr$,
the quantities $q_k^{(n,pn)}$ are polynomially expressed with the help of
(\ref{r}) upon $\rho_k\equiv q_k^{(n,r)}$.

Let us exhibit some lattices which can be found in the literature\footnote{Denote
$\phantom{q}^{\prime}=\partial/\partial t_p^{(n)}$ with some corresponding
values of $p$ and $n$}.

{\sc Example 1.}
$n\geq 2,\: p=1,\: r=1,\: l=1,$ \cite{na1}, \cite{bo1}
\begin{equation}
\rho_i^{\prime} =
\rho_i\left(\sum_{s=1}^{n-1}\rho_{i+s}  -
\sum_{s=1}^{n-1}\rho_{i-s}\right),\;\;
\rho_1(i)\equiv\rho_i,
\label{b}
\end{equation}

{\sc Example 2.}
$n=p+1,\: p\geq 1,\: r=p,\: l=1,$ \cite{bo1}
\[
\rho_i^{\prime} = \rho_i\left(\prod_{s=1}^{n-1}\rho_{i+s}  -
\prod_{s=1}^{n-1}\rho_{i-s}\right),
\]

{\sc Example 3.}
$n=1,\: p=1,\: r=1,\: l\geq 2,$ \cite{ku1}
\begin{equation}
\begin{array}{c}
\rho_1^{\prime}(i) = \rho_2(i+1) - \rho_2(i),\\[0.5cm]
\rho_k^{\prime}(i) = \rho_{k+1}(i+1) - \rho_{k+1}(i) +
\rho_k(i)\left(\rho_1(i) - \rho_1(i-k+1)\right),\: k=2,...,l-1,\\[0.5cm]
\rho_l^{\prime}(i) = \rho_l(i)
\left(\rho_1(i) - \rho_1(i-l+1)\right),
\end{array}
\label{kuper}
\end{equation}

{\sc Example 4.}
$n=1,\: p=r,\: r\geq 1,\: l=p+1,$ \cite{bl1}
\[
\rho_1^{\prime}(i) = \rho_{p+1}(i+p) - \rho_{p+1}(i),
\]
\[
\rho_k^{\prime}(i) = \rho_{k-1}(i)\rho_{p+1}(i+p-k+1) -
\rho_{k-1}(i-1)\rho_{p+1}(i),\: k=2,...,p+1,
\]

{\sc Example 5.}
$n\geq 1,\: p=1,\: r=-1,\: l=2$
\[
\rho_1^{\prime}(i) = \rho_{2}(i+n) - \rho_{2}(i) +
\rho_1(i)\left(\sum_{s=1}^{n}\rho_1(i-s)-\sum_{s=1}^{n}\rho_1(i+s)\right),
\]
\begin{equation}
\rho_2^{\prime}(i) = \rho_2(i)\left(\sum_{s=1}^{n}\rho_1(i-s-n) -
\sum_{s=1}^{n}\rho_1(i+s)\right).
\label{belov}
\end{equation}

The system (\ref{b}) is known as Bogoyavlenskii lattice (cf. \cite{na1}).
In particular case $n=2$, this is Volterra lattice.
The system (\ref{kuper}) is known in the literature as generalized
Toda lattice or Kupershmidt lattice. In particular case $n=1$,
we have ordinary Toda lattice.
The system (\ref{belov}), for $n=1$, was considered in \cite{be1}.
All these examples and many others can be found in \cite{ci1}.

{\sc Example 6.}
$n=2,\: p=1,\: r=3,\: l=2$ \cite{hu1}
\[
(\rho_1(i-1) + \rho_1(i) + \rho_1(i+1))^{\prime} =
\]
\[
= (\rho_1(i-1) + \rho_1(i) + \rho_1(i+1))\left(\rho_1(i-1)-\rho_1(i+1)\right) +
\rho_2(i+1) - \rho_2(i-1),
\]
\begin{equation}
\rho_2^{\prime}(i) = \rho_2(i)\left(\rho_1(i+1) - \rho_1(i-1)\right).
\label{hu}
\end{equation}
Здесь $\rho_1(i)\equiv q_1^{(2,1)}(i),\: \rho_2(i)\equiv q_2^{(2,3)}(i)$.

{\sc Example 7.}
$n=1,\: p=1,\: r\geq 2,\: l=r$
\[
(\rho_1(i) + ... + \rho_1(i+r-1))^{\prime} =
\]
\[
=(\rho_1(i) + ... + \rho_1(i+r-1))\left(\rho_1(i)-\rho_1(i+r-1)\right) +
\rho_2(i+1) - \rho_2(i),
\]
\begin{equation}
\rho_k^{\prime}(i) = \rho_k(i)\left(\rho_1(i) - \rho_1(i+r-k)\right) +
\rho_{k+1}(i+1) - \rho_{k+1}(i),\;\;
k = 2,..., r-2,
\label{sh}
\end{equation}
\[
\rho_{r-1}^{\prime}(i) = \rho_{r-1}(i)\left(\rho_1(i) - \rho_1(i+1)\right)
+\mu_{i+1} - \mu_{i}.
\]
Here $\rho_1(i)\equiv q_1^{(1,1)}(i),\: \rho_k(i)\equiv q_k^{(1,r)}(i)$;
$\mu_i = q_r^{(1,r)}(i)$ are constants by virtue of motion equations.
The system (\ref{sh}) describes Darboux transformation of purely
differential operator of $r$-th order \cite{ad1}. Indeed, due to
\cite{sv2} one can write
\[
{\cal Q}_i^l = H_{i+(l-1)n}...H_{i+n}H_i +
q_1^{(n, ln)}(i)H_{i+(l-2)n}...H_{i+n}H_i + ...
\]
\[
+ q_l^{(n, ln)}(i) +
q_{l+1}^{(n, ln)}(i)H_{i-n}^{-1} +
q_{l+2}^{(n, ln)}(i)H_{i-2n}^{-1}H_{i-n}^{-1} + ...,
\]
where $H_i\equiv \partial^{(n)} - q_1^{(n,n)}(i)$ is the operator defining
elementary Darboux transformation: $z\psi_{i+n} = H_i\psi_i$.
Then on ${\cal S}_{1,r,r}$ we have
\[
{\cal Q}_i^r = H_{i+r-1}...H_{i+1}H_i +
\rho_1^{(r)}(i)H_{i+r-2}...H_{i+1}H_i +
\rho_2(i)H_{i+r-3}...H_{i+1}H_i + ... + \mu_i,
\]
where  $\rho_1^{(r)}(i)\equiv \rho_1(i) + ... + \rho_1(i+r-1)$. As is known,
the condition ${\cal Q}_i^r = ({\cal Q}_i^r)_{+}$ defines reduction
of KP hierarchy to Gelfand-Dickey one.

{\sc Remark 6.}
The systems (\ref{hu}) and (\ref{sh}), in contrast with above-mentioned
examples do not admit the representation (\ref{ustinov}),
since in this case $pn/r$'s are ratios.

The rest of this section will be concerned with constructing of some
class of one-component lattices which naturally contains Bogoyavlenskii lattices.
For simplicity, let us consider the system which describes
the evolution on ${\cal S}_{4,2,1}$.
For $\rho = q_1^{(4,2)}$ we have
\[
\partial^{(4)}\rho_i = \rho_i(\rho_{i+2} - \rho_{i-2}).
\]
This is Volterra lattice with double spacing. From (\ref{r}) we have
$\rho_i = \sigma_i + \sigma_{i+1}$, where
$\sigma\equiv q_1^{(4,1)}$.
Moreover, on ${\cal S}_{4,2,1}$, we have
\[
q_2^{(4,2)}(i) = 0 =  q_2^{(4,1)}(i+1) + q_2^{(4,1)}(i) +
q_1^{(4,1)}(i-3)q_1^{(4,1)}(i)
\]
or
\[
q_2^{(4,1)}(i+1) + q_2^{(4,1)}(i) = -\sigma_{i-3}\sigma_i.
\]
Then
\begin{equation}
q_2^{(4,1)}(i+4) - q_2^{(4,1)}(i) =  \sigma_{i-3}\sigma_i -
\sigma_{i-2}\sigma_{i+1} + \sigma_{i-1}\sigma_{i+2} - \sigma_i\sigma_{i+3}.
\label{differ}
\end{equation}
It follows from (\ref{expl.form}) that
\[
\partial^{(4)}\sigma_i = \partial^{(4)}q_1^{(4,1)}(i) =
q_2^{(4,1)}(i+4) - q_2^{(4,1)}(i) + q_1^{(4,1)}(i)(q_1^{(4,4)}(i) -
q_1^{(4,4)}(i-3)).
\]
Taking into account $q_1^{(4,4)}(i) = \sigma_i + \sigma_{i+1} + \sigma_{i+2} +
\sigma_{i+3}$
and the relation (\ref{differ}), as a result, we gets
\[
\sigma_{i}^{\prime} = \sigma_{i}(\sigma_{i+2} + \sigma_{i+1} -
\sigma_{i-1} -
\sigma_{i-2}) +
\sigma_{i+2}\sigma_{i-1} - \sigma_{i-2}\sigma_{i+1}.
\]

Now let us generalize this example. On ${\cal S}_{mr,r,1}$, where
$m\geq 2$ и $r\geq 1$, we have
\[
\partial^{(mr)}\rho_i =
\rho_i\left(\sum_{s=1}^{m-1}\rho_{i+sr}  -
\sum_{s=1}^{m-1}\rho_{i-sr}\right).
\]
Here $\rho = q_1^{(mr,r)}$.
Let $\sigma = q_1^{(mr,1)}$, then
$\rho_i = \sigma_i + ... + \sigma_{i+r-1}$. Let us, omitting technical
details, to write down the equations on the field $\sigma$
\[
\sigma_i^{\prime} = \sigma_i\left(\sum_{s=1}^{(m-1)r}\sigma_{i+s} -
\sum_{s=1}^{(m-1)r}\sigma_{i-s}\right) +
\]
\[
+ \sum_{s=1}^{m-1}\sigma_{i+sr}\left(\sum_{s_1=1}^{r-1}\sigma_{i-s_1+(s-m+1)}
\right) -
\sum_{s=1}^{m-1}\sigma_{i-sr}\left(\sum_{s_1=1}^{r-1}\sigma_{i+s_1-(s-m+1)}
\right).
\]

\noindent
{\bf 4.2. The relationship of constrained KP hierarchies with integrable
lattices.}

As is known, integrable lattices are treated as discrete symmetries
(Darboux transformations) for corresponding integrable (differential)
hierarchies.

{\sc Proposition 4.}
{\it Let the solution of the DKP chain is in  ${\cal S}_{n,r,l}$. Denote
$m=ln - r$,
\[
\psi = \psi_i,\;\;
\tilde{\psi} = \psi_{i+nm},\;\;
a = z\frac{\tilde{\psi}}{\psi} =
z\frac{\psi_{i+nm}}{\psi_i} = z^{1-nm}a^{[nm]}(i).
\]
1) If $m\geq 0$, then the triple $(h, \tilde{h}, a)$
is a solution of DKP hierarchy such that
\begin{equation}
z^{\ell}a\in{\cal H}_{+},\; z^{\ell + r}a\in{\cal H}_{+},
\label{q1}
\end{equation}
where $\ell = m-1\geq -1$.

2) If $m<0$, then the triple $(h, \tilde{h}, a)$ is a solution
of DKP hierarchy satisfiyng
\begin{equation}
z^{-\ell}a^{-1}\in\tilde{{\cal H}}_{+},\; z^{\ell + r}a\in{\cal H}_{+}.
\label{q2}
\end{equation}
}

{\sc Proof.} We prove the first part of proposition. The second one
can be proved by analogy. By virtue of Proposition 1 and its corollary
$(h, \tilde{h}, a)$ is indeed a solution of DKP hierarchy. In the
circumstances of this proposition we have
\begin{equation}
z^{1-n}a^{[n]}(i)\in{\cal H}_{+}(i),\;
z^{l(1-n)+r}a^{[ln-r]}\in{\cal H}_{+}(i).
\label{q3}
\end{equation}
Using the reasonings taking into account
when proving  Proposition 2, as consequence of  (\ref{q3}), we obtain
\[
z^{(1-n)m}a^{[nm]}(i)\in{\cal H}_{+}(i),\;
z^{(l-m)n}a^{[mn]}\in{\cal H}_{+}(i).
\]
These relations, in turn, can be rewritten in the form (\ref{q1}).

As was mentioned above, the relations (\ref{q1}) determine Krichever's
rational reductions of KP hierarchy including (for $\ell = -1$) Gelfand-Dickey
ones. In this situation $r$-th power of Lax operator is expressed in the form
of ratio of two differential operators (\ref{rat}). Using standard reasonings,
one can derive that from (\ref{q2}) it follows, that
\[
{\cal Q}^r = P_{|\ell + 1|}Q_{\ell + r + 1}.
\]

It is worthwhile to note some works in which the relationship between 
integrable lattices and constrained KP hierarchies is mentioned. The article
\cite{ar1}  is conserned with discrete symmetries for the so-called 
multi-boson hierarchies with Lax operator of the form
\[
{\cal Q} = \partial +
\sum_{k=1}^nR_k(\partial - S_k)^{-1}...(\partial - S_2)^{-1}(\partial - S_1)^{-1}.
\]
These symmetries is given in its explicit form as shifts on generalized
Toda lattices (\ref{kuper}). In \cite{sv3} it was constructed the 
modified version of Krichever's rational reductions. The approach in \cite{sv3}
essentialy uses some class of one-component lattices
(${\cal S}_{n,r,1},\; r = 1, ..., n-1$). In \cite{sv3} also the discrete 
symmetries are constructed. The results of these two works as can be shown
are compatible between each other and are in agreement with the Proposition 4.

It should be noted here that the relationship between discrete integrable
systems and restricted KP hierarchies is considerably used in  matrix models
(see, for example \cite{ar2}, \cite{bo2}).

\noindent
{\bf 4.2. Lattice Miura transformations.}

Given any solution of DKP chain define, for some $k\in\dZ^{*}$,
the following transformation:
\begin{equation}
g_k : \left\{
\begin{array}{l}
a(i)\rightarrow z^{1-k}a^{[k]}(ki), \\
h(i)\rightarrow h(ki).
\end{array}
\right.
\label{trans}
\end{equation}

{\sc Lemma.}
{\it The relation
\begin{equation}
\overline{a}^{[r]}(i) = z^{r(1-k)}a^{[rk]}(ki),\;\;
\forall\: r, k\in\dZ^{*},
\label{tr}
\end{equation}
is valid where $\overline{a}(i)\equiv g_k(a(i))$.
}

{\sc Proof.} For $r>0$, we have
\[
\overline{a}^{[r]}(i)\equiv\overline{a}(i)\overline{a}(i+1)...\overline{a}(i+r-1)
= z^{r(1-k)}a^{[k]}(ki)a^{[k]}(ki+k)...a^{[k]}(ki+(r-1)k) =
\]
\[
= z^{r(1-k)}a^{[rk]}(ki),
\]
while for $r<0$,
\[
\overline{a}^{[r]}(i)\equiv
\overline{a}^{-1}(i-1)\overline{a}^{-1}(i-2)...\overline{a}^{-1}(i-|r|) =
\]
\[
= z^{|r|(k-1)}a^{[-k]}(ki)a^{[-k]}(ki-k)...a^{[k]}(ki-(|r|-1)k)
= z^{r(1-k)}a^{[rk]}(ki).
\]
In the latter case we use easily checked identity
\[
\overline{a}^{-1}(i-1) = z^{k-1}a^{[-k]}(ki).
\]

{\sc Proposition 5.}
{\it The set of transformation (\ref{trans}) with superposition operation
is isomorphic to multiplicative semi-group $\dZ^{*}$.}

{\sc Proof.}
By virtue (\ref{tr}), we have
\[
g_r\circ g_k(a(i)) = z^{1-r}\overline{a}^{[r]}(ri) =
\]
\[
= z^{1-r}z^{r(1-k)}a^{[rk]}(rki) = z^{1-rk}a^{[rk]}(rki) = g_{rk}(a(i)).
\]

{\sc Proposition 6.}
{\it (\ref{trans}) is symmetry transformation for DKP chain.
}

{\sc Proof.}
By virtue of Proposition 1, we have
\[
\partial_p\overline{a}(i) = z^{1-k}\partial_pa^{[k]}(ki) =
\]
\[
= z^{1-k}a^{[k]}(ki)(H^{(p)}(ki+k) - H^{(p)}(ki)) =
\overline{a}(i)(\overline{H}^{(p)}(i+1) - \overline{H}^{(p)}(i)).
\]
In addition,
\[
\partial_p\overline{h}(i) = \partial_ph(ki) = \partial H^{(p)}(ki) =
\partial\overline{H}^{(p)}(i).
\]

{\sc Proposition 7.}
{\it Let $\{h(i), a(i)\}\in {\cal S}_l^{kn}$, then
$\{g_k(h(i)), g_k(a(i))\}\in {\cal S}_l^{n}$
}

{\sc Proof.} If $z^{l-kn+1}a^{[kn]}(i)\in{\cal H}_{+}(i)$, then
by virtue (\ref{tr}) we have
\[
z^{l-n+1}\overline{a}(i) = z^{l-n+1}z^{n(1-k)}a^{[kn]}(ki)
= z^{l-kn+1}a^{[kn]}(ki)\in{\cal H}_{+}(ki) = \overline{{\cal H}}_{+}(i).
\]
So, we can write $g_k({\cal S}_l^{kn})\subset {\cal S}_l^n$.

It is natural that the transformation (\ref{trans}) affects corresponding
transformation of the functions $q_k^{(n,r)}(i)$. To find out the rule of
how they transform we use (\ref{REL}). By virtue of (\ref{tr}), we
have
\[
z^r =
\overline{a}^{[r]}(i) +
\sum_{s\geq 1}\overline{q}_s^{(n,r)}(i)z^{s(n-1)}\overline{a}^{[r-sn]}(i) =
\]
\[
= z^{r(1-k)}a^{[rk]}(ki) +
\sum_{s\geq 1}\overline{q}_s^{(n,r)}(i)z^{s(n-1)}z^{(r-sn)(1-k)}
a^{[(r-sn)k]}(ki).
\]
Multiplying both sides of this relation by $z^{r(k-1)}$, we obtain
\[
z^{rk} =
a^{[rk]}(ki) +
\sum_{s\geq 1}\overline{q}_s^{(n,r)}(i)z^{s(kn-1)}a^{[kr-skn]}(ki).
\]
It follows from this following identification
\begin{equation}
\overline{q}_s^{(n,r)}(i) = q_s^{(kn,kr)}(ki).
\label{miura}
\end{equation}
Observe, that if one wants to get transformation in the form of the mapping
$\{q_s^{(kn,r)}(i)\}\rightarrow
\{\overline{q}_s^{(n,r)}(i)\}$, there is a need to make use the formula (\ref{r}).
As an example, take $n=1, r=1, k=2$. Let
$\overline{q}_s(i)\equiv\overline{q}_s^{(1,1)}(i)$ and
$q_s(i)\equiv q_s^{(2,1)}(i)$, then
\[
\overline{q}_1(i) = q_1(2i) + q_1(2i+1),\;\;
\overline{q}_2(i) = q_2(2i) + q_1(2i-1)q_1(2i) + q_1(2i+1),\;\;
\]
\[
\overline{q}_3(i) = q_3(2i) + q_1(2i)q_2(2i-1) + q_1(2i-3)q_2(2i) + q_3(2i+1)
\]
and so on. These relation serve to map the solutions of gap KP hierarchy,
for $n=2$, to solutions of discrete KP hierarchy. In particular, if
$q_s(i)\equiv 0$ $s\geq 2$, one gets well-known Miura transformation
between Volterra and Toda lattices
\begin{equation}
\overline{q}_1(i) = q_1(2i) + q_1(2i+1),\;\;
\overline{q}_2(i) = q_1(2i-1)q_1(2i).
\label{MT}
\end{equation}
Observe, that $g_k$ only for $k=-1$ and $k=1$ is invertible transformation.
It is natural, for $k\neq\pm 1$, (\ref{trans}) (or (\ref{miura}) to cal
lattice Miura transformation.

{\sc Proposition 8.}
{\it Let the number $kln-r$ do not multiply by $k$; the DKP chain solution
is in ${\cal S}_0^{kn}\cap {\cal S}_{l-1}^{kln-r}$;
the submanifolds ${\cal S}_0^{kn}$ and
$S_{l-1}^{kln-r}$ are origins of inclusions chains for given solution, then
$g_k({\cal S}_0^{kn}\cap {\cal S}_{l-1}^{kln-r})\subset {\cal S}_0^{n}\cap
{\cal S}_{kl-1}^{kln-r}$.
Moreover  ${\cal S}_0^n$ and ${\cal S}_{kl-1}^{kln-r}$
are origins of inclusions chains for transformed solution.
}

{\sc Proof.} By condition, we have two chains of submanifolds
\[
{\cal S}_0^{kn}\subset{\cal S}_1^{2kn}\subset ...
\]
\[
{\cal S}_{l-1}^{lkn-r}\subset{\cal S}_{2l-1}^{2(lkn-r)}\subset ...\subset
{\cal S}_{kl-1}^{k(lkn-r)}\subset...
\]
consisting given solution of DKP chain.
By virtue of Proposition 7, transformed solution is in
submanifolds
\[
{\cal S}_0^{n}\subset{\cal S}_1^{2n}\subset ...
\]
\[
{\cal S}_{kl-1}^{lkn-r}\subset{\cal S}_{2kl-1}^{2(lkn-r)}\subset...
\]
So, the proposition is proved.

By virtue of this proposition one can write
\[
g_k({\cal S}_{kn,r,l})\subset{\cal S}_{n,r,kl}.
\]
Together with (\ref{miura}) Proposition 8 is a strong basis for constructing
of Miura transformations between lattices with finite number of fields.
Simplest example is given by (\ref{MT}). We learn, for example,
from Proposition 8,  that Bogoyavlenskii lattices (\ref{b}) are connected
by Miura transformations with generalized Toda ones (\ref{kuper}).
In this case one can write $g_n({\cal S}_{n,1,1})\subset
{\cal S}_{1,1,n},\; n\geq 2$.

Some examples of lattice Miura transformations which connect one-component
lattices to multi-component ones can be found in \cite{sv2},
\cite{sv3}.

%---------------------------------------------------------------------------

\section{Self-similar solutions}

\noindent
{\bf 5.1. Invariant solutions.}

It is evident, that linear systems  (\ref{evolution}), (\ref{eigenv.pr.})
and its consistency relations (\ref{expl.form}) and (\ref{r})
are invariant under group of dilatations
\[
q_k^{(n,r)}(i)\rightarrow\epsilon^k q_k^{(n,r)}(i),\;\;
t_l\rightarrow\epsilon^{-l}t_l,\;\;
z\rightarrow\epsilon z,\;\;
\Psi_i\rightarrow\epsilon^i\Psi_i.
\]
In what follows we consider dependencies only on finite number of evolution
parameters $t_1,..., t_p$. Invariants of this group are
\[
T_l = \frac{t_l}{(pt_p)^{l/p}},\;
l = 1,..., p-1,\;\;
\xi = (pt_p)^{1/p}z,\;\;
\psi_i = z^i\Psi_i.
\]
From this we gets the ansatzes for self-similar solutions:
\begin{equation}
q_k^{(n,r)}(i) = \frac{1}{(pt_p)^{k/p}}x_k^{(n,r)}(i),
\label{substitution}
\end{equation}
\[
\Psi_i = z^i\psi_i(\xi; T_1,..., T_{p-1}).
\]
Here $x_k^{(n,r)}(i)$'s are unknown functions of $T_1,..., T_{p-1}$.
Direct substitution of (\ref{substitution}) into (\ref{expl.form}) gives
\begin{equation}
\partial_{T_l}x_k^{(n,r)}(i) = X_{k,l}^{(n,r)}(i),\;\;
l =1,..., p-1
\label{ev}
\end{equation}
and
\[
kx_k^{(n,r)}(i) + T_1X_{k,1}^{(n,r)}(i) + 2T_2X_{k,2}^{(n,r)}(i) + ...
+ (p-1)T_{p-1}X_{k,p-1}^{(n,r)}(i) +
\]
\begin{equation}
+ X_{k,p}^{(n,r)}(i) = 0,\;\;
l =1,..., p-1.
\label{ev1}
\end{equation}
Here $X^{(n,r)}_{k,l}(i)$'s are RHS's of (\ref{expl.form}) where
$q^{(n,r)}_k(i)$'s are replaced by  $x^{(n,r)}_k(i)$'s.
Moreover it follows from (\ref{r}) that
\[
x_k^{(n,r_1+r_2)}(i) =
x_k^{(n,r_1)}(i) +
\sum_{s=1}^{k-1}x_s^{(n,r_1)}(i)x_{k-s}^{(n,r_2)}(i+r_1-sn)
+ q_k^{(n,r_2)}(i+r_1) =
\]
\begin{equation}
= x_k^{(n,r_2)}(i)
+ \sum_{s=1}^{k-1}x_s^{(n,r_2)}(i)x_{k-s}^{(n,r_1)}(i+r_2-sn)
+ x_k^{(n,r_1)}(i+r_2).
\label{r1}
\end{equation}
Corresponding auxiliary linear equations are transformed
to the following form:
\[
\partial_{T_l}\psi = (X_{(n)}^{ln})_{+}\psi,\;\;
l = 1,..., p-1,
\]
\[
\xi\psi_{\xi} = \left\{T_1(X_{(n)}^n)_{+} + 2T_2(X_{(n)}^{2n})_{+} + ... +
(p-1)(X_{(n)}^{(p-1)n})_{+} + (X_{(n)}^{pn})_{+}\right\}\psi,
\]
\[
X_{(n)}^r\psi = \xi\psi,
\]
where
\[
X_{(n)}^r\equiv\xi\Lambda^r + \sum_{k\geq 1}\xi^{1-k}x_k^{(n,r)}\Lambda^{r-kn},\;\;
r\in\dZ.
\]

{\bf 5.2. Examples.}

Let us consider, as a simple example, the case corresponding
Volterra lattice hierarchy, that is $n=2, r=1, l=1$. Take $p=2$. The equations
(\ref{ev}) and (\ref{ev1}) are written down as follows:
\begin{equation}
x_i^{\prime} = x_i(x_{i+1} - x_{i-1}),\;\;
\phantom{q}^{\prime}\equiv\partial/\partial T_1,
\label{ev2}
\end{equation}
\begin{equation}
x_i + T_1x_i\left\{x_1^{(2,2)}(i) - x_1^{(2,2)}(i-1)\right\} +
x_i\left\{x_2^{(2,4)}(i) - x_2^{(2,4)}(i-1)\right\} = 0.
\label{ev3}
\end{equation}
Here we denote $x_i = x_1^{(2,1)}(i)$. Using (\ref{r1}) one calculates
\[
x_1^{(2,2)}(i) = x_i + x_{i+1},\;\;
x_2^{(2,4)}(i) = x_i(x_{i-1} + x_i + x_{i+1})
+ x_{i+1}(x_i + x_{i+1} + x_{i+2}).
\]
Taking into account these relations, the equation (\ref{ev3}) turns into
\begin{equation}
\begin{array}{l}
T_1x_i + x_i(x_{i-1} + x_i + x_{i+1}) = \alpha_i, \\
1+\alpha_{i+1}-\alpha_{i-1} = 0.
\end{array}
\label{dp1}
\end{equation}
Let us prove that  by virtue of (\ref{ev2}) $\alpha_i$'s constants.
Indeed, we have
\[
\alpha_i^{\prime} = x_i + T_1x_i(x_{i+1} - x_{i-1}) +
\]
\[
+ x_i(x_{i+1} - x_{i-1})(x_{i-1} + x_i + x_{i+1}) +
x_i(x_{i+1}x_{i+2} - x_{i-1}x_{i-2}) =
\]
\[
= x_i(1+\alpha_{i+1}-\alpha_{i-1}) = 0.
\]
So, one can rewrite (\ref{dp1}) as
\begin{equation}
x_{i-1} + x_i + x_{i+1} = -T_1 + \frac{\alpha_i}{x_i}.
\label{dp2}
\end{equation}
Here $\alpha_i$'s are constants forced by the condition
$\alpha_{i+2} = \alpha_i - 1$. One can write the solution of this equation
as $\alpha_i = \alpha - \frac{1}{2}i + \beta(-1)^i$ where
$\alpha$ and $\beta$ are some constants. Provided these conditions,
(\ref{dp2}) is dPI \cite{gr1}.

Observe that evolution equation (\ref{ev2}) with (\ref{dp2}) turns into
\[
x_i^{\prime} = 2x_ix_{i+1} + x_i^2 + T_1x_i - \alpha_i.
\]
It can be easily checked that together with (\ref{dp2}) this lattice
is equivalent to the pair of ordinary first-order differential equations
\begin{equation}
\begin{array}{l}
w_1^{\prime} = 2w_1w_2 + w_1^2 + T_1w_1 + a, \\
w_2^{\prime} = -2w_1w_2 - w_2^2 - T_1w_2 - b
\end{array}
\label{twin}
\end{equation}
with discrete symmetry transformation
\[
\overline{w}_1 = w_2,\;\;
\overline{w}_2 = - w_1 - w_2 - T_1 -\frac{b}{w_2},\;\;
\overline{a} = b,\;\;
\overline{b} = a + 1,
\]
where $w_1\equiv x_i,\; w_2 = x_{i+1},\; a\equiv -\alpha_i,\;
b\equiv -\alpha_{i+1}$ for some fixed (but arbitrary)
value $i=i_0$. In turn  the system (\ref{twin}) is equivalent to second-order
equation
\[
w^{\prime\prime} = \frac{(w^{\prime})^2}{2w} + \frac{3}{2}w^3 + 2T_1w^2 +
\left(\frac{T_1^2}{2} + a - 2b + 1\right)w - \frac{a^2}{2w},\;\;
w\equiv w_1
\]
with corresponding symmetry transformation
\[
\overline{w} =
\frac{w^{\prime} - w^2 - T_1w - a}{2w},\;\;
\overline{a} = b,\;\;
\overline{b} = a + 1.
\]
In fact this is PIV with B\"acklund transformation
\cite{lu1}, \cite{gr2}.
By dilatations
$T_1\rightarrow\sqrt{2}T_1,\; w\rightarrow w/\sqrt{2}$ it can be turned
to following canonical form:
\begin{equation}
w^{\prime\prime} = \frac{(w^{\prime})^2}{2w} + \frac{3}{2}w^3 + 4T_1w^2 +
2(T_1^2 + a - 2b + 1)w - \frac{2a^2}{w}
\label{P42}
\end{equation}
\[
\overline{w} =
\frac{w^{\prime} - w^2 - 2T_1w - 2a}{2w},\;\;
\overline{a} = b,\;\;
\overline{b} = a + 1.
\]
In a result, we obtain the well-known relationship between dPI (\ref{dp2})
and PIV (\ref{P42}) \cite{gr3}.

{\sc Remark 7.}
The equations (\ref{twin}) can be interpreted as self-similar reduction
of Levi system \cite{mar}
\[
\begin{array}{l}
v_{1t_2} = (-v_1^{\prime} + v_1^2 + 2v_1v_2)^{\prime}, \\
v_{2t_2} = (v_2^{\prime} + v_2^2 + 2v_1v_2)^{\prime}
\end{array}
\]
with the help of the ansatze
\[
v_k=\frac{1}{(2t_2)^{1/2}}w_k(T_1),\;\;
k = 1, 2.
\]
This fact is suitable for results of the work \cite{sv3},
where we established the correspondence between some class of one-component
lattices and hierarchies of evolution equations which can be interpreted
as modified version of Krichever's rational reductions of KP hierarchy
In particular, Levi system hierarchy corresponds to Volterra lattice.

Let us consider more general case corresponding to Bogoyavlenskii lattices
(\ref{b}). The analoges of the equations (\ref{ev2}) and (\ref{ev3}) in this
case are Volterra lattice.
\begin{equation}
x_i^{\prime} = x_i\left(\sum_{s=1}^{n-1}x_{i+s} -
\sum_{s=1}^{n-1}x_{i-s}\right),
\label{bv2}
\end{equation}
\begin{equation}
x_i + T_1x_i\left\{x_1^{(n,n)}(i) - x_1^{(n,n)}(i+1-n)\right\} +
x_i\left\{x_2^{(n,2n)}(i) - x_2^{(n,2n)}(i+1-n)\right\} = 0.
\label{v3}
\end{equation}
Moreover we take into account that
\[
x_1^{(n,n)}(i) = \sum_{s=1}^n x_{i+s-1},\;\;
x_2^{(n,2n)}(i) = \sum_{s=1}^n x_{i+s-1}\left(
\sum_{s_1=1}^{2n-1}x_{i+s_1+s-n-1}\right).
\]
Then the equation (\ref{v3}) is rewritten in the form
\[
T_1x_i + x_i(x_{i+1-n} + ... +  x_{i+n-1}) = \alpha_i,
\]
\begin{equation}
1 + \sum_{s=1}^{n-1}\alpha_{i+s} - \sum_{s=1}^{n-1}\alpha_{i-s} = 0.
\label{2}
\end{equation}
It can be proved, that by virtue of (\ref{bv2}),
\[
\alpha_i^{\prime}
= x_i\left(1+\sum_{s=1}^{n-1}\alpha_{i+s}-\sum_{s=1}^{n-1}\alpha_{i+s}\right)
= 0.
\]
So, one concludes that in this case self-similar ansatze leads to
equation
\begin{equation}
x_{i+1-n} + ... + x_{i+n-1} = -T_1 + \frac{\alpha_i}{x_i},
\label{dp3}
\end{equation}
when the constants $\alpha_i$ is connected with each other by (\ref{2}).

Standard analysis of singularity confinement shows that this property for
(\ref{dp3}) is valid provided that
\begin{equation}
\alpha_{i+n} -
\alpha_i = \alpha_{i+2n-1} - \alpha_{i+n-1}.
\label{usl}
\end{equation}
This equation do not contradict to (\ref{2}), but is more general.
Let us show it.
It follows from (\ref{2}) that
\begin{equation}
-1 = \sum_{s=1}^n\alpha_{i+s+n-1} - \sum_{s=1}^n\alpha_{i+s} =
\sum_{s=1}^n\alpha_{i+s+n-2} - \sum_{s=1}^n\alpha_{i+s-1}.
\label{usl1}
\end{equation}
Second equality in (\ref{usl1}) can be rewritten in the form
\[
\sum_{s=2}^{n-1}\alpha_{i+s+n-1} + \alpha_{i+2n-1}
- \sum_{s=1}^{n-2}\alpha_{i+s} - \alpha_{i+n-1} =
\alpha_{i+n} + \sum_{s=3}^{n}\alpha_{i+s+n-2}
- \alpha_i - \sum_{s=2}^{n-1}\alpha_{i+s-1}.
\]
From the latter one obtains (\ref{usl}).

As is known, Bogoyavlenskii lattice (\ref{b}), for any $n$,
can be interpreted as integrable discretization of Korteweg-de Vries
equation. Similarly, autonomous version of (\ref{dp3}), for any $n$,
is integrable discretization of PI: $w^{\prime\prime} = 6w^2 + t$.
Let us prove this. Let $\alpha_i=\alpha$. One divides real axis into
segments of equal length $\varepsilon$. So, it can be written $t=i\varepsilon$.
Values of the function $w$, respectively, is taken for all such values of the
variable $t$. Then one can denote $w(t) = w_i$. Let
\begin{equation}
x_i = 1 + \varepsilon^2w_i,\;\;
\alpha = 1 - 2n - \varepsilon^4t,\;\;
T_1 = -(2n-1).
\label{s}
\end{equation}
Substituting (\ref{s}) in the equation (\ref{dp3}), taking into
account the relations of the form
\[
x_{i+1} = 1 + \varepsilon^2w_{i+1} = 1 + \varepsilon^2\{w +
\varepsilon w^{\prime} + \frac{\varepsilon^2}{2}w^{\prime\prime} + ...\}
\]
and turning then $\varepsilon$ to zero we obtain, in continuous limit
the equation
\[
\sum_{s=1}^{n-1}(n-s)^2\cdot w^{\prime\prime} = - t - (2n-1)w^2
\]
which by dilatations can be deduced to canonical form of PI.

To conclude the paper, we exhibit two examples of integrable mappings
and suitable Miura transformations.

{\sc Example 8.}
Two-component mapping
\[
\beta_i = T_1y_1(i) + y_1^2(i) + y_2(i) + y_2(i+1),
\]
\begin{equation}
y_1(i) +
y_2(i+1)(y_1(i) + y_1(i+1) + T_1)
- y_2(i)(y_1(i) + y_1(i-1) + T_1) = 0,
\label{ex1}
\end{equation}
\[
2 + \beta_i - \beta_{i-1} = 0.
\]
is suitable for Toda lattice hierarchy.
The connection with dPI (\ref{dp1}) is given
by Miura transformation
\begin{equation}
y_1(i) = x_{2i} + x_{2i+1},\;\;
y_2(i) = x_{2i-1}x_{2i},\;\;
\beta_i = \alpha_{2i} + \alpha_{2i+1}.
\label{m1}
\end{equation}
For example, the substitution of (\ref{m1}) into (\ref{ex1}) gives
\[
x_{2i}(1+\alpha_{2i+1}-\alpha_{2i-1}) +
x_{2i+1}(1+\alpha_{2i+2}-\alpha_{2i}) = 0.
\]

{\sc Example 9.}
Discrete equations
\[
\beta_i = T_1y_1(i) + y_1^2(i) + y_2(i) + y_2(i+1),
\]
\begin{equation}
y_1(i) +
y_2(i+1)(y_1(i) + y_1(i+1) + T_1)
- y_2(i)(y_1(i) + y_1(i-1) + T_1) + y_3(i+2) - y_3(i) = 0,
\label{ex2}
\end{equation}
\[
y_2(i)(2 + \beta_i - \beta_{i-1}) +
y_3(i+1)(y_1(i) + y_1(i+1) + T_1) -
\]
\[
- y_3(i)(y_1(i-1) + y_1(i-2) + T_1) = 0,
\]
\[
3 + \beta_i - \beta_{i-2} = 0.
\]
correspond to three-component generalized Toda lattice (\ref{kuper}) hierarchy.
The system (\ref{ex2}) is connected by Miura transformation
\[
y_1(i) = x_{3i} + x_{3i+1} + x_{3i+2},\;\;
y_2(i) = x_{3i-2}x_{3i} + x_{3i-1}x_{3i} + x_{3i-1}x_{3i+1},\;\;
y_3(i) = x_{3i-4}x_{3i-2}x_{3i},\;\;
\]
\[
\beta_i = \alpha_{3i} + \alpha_{3i+1} + \alpha_{3i+2}
\]
with (\ref{dp3}), for $n=3$.

{\bf Acknowledgments.} This work was supported by grants RFBR 03-01-00102
and INTAS 2000-15.

\end{document}